# BUNCH LENGTHENING AND MICROWAVE INSTABILITY IN THE DAΦNE POSITRON RING

*M. Zobov, A. Drago, A. Gallo, A. Ghigo, F. Marcellini,*
*M. Migliorati, L. Palumbo, M. Serio, G. Vignola*

## 1. Introduction

We have performed a comprehensive study of the bunch lengthening and microwave instability in the DAΦNE main rings which includes analytical estimates, numerical simulations and measurements.

First, we have applied a double Water Bag distribution model and solved the Vlasov equation in order to investigate the bunch longitudinal coherent mode coupling leading to the microwave instability and evaluate the instability threshold.

Then, very time consuming numerical calculations have been performed to simulate the bunch lengthening process and estimate the energy spread growth (bunch widening) beyond the microwave threshold. For this purpose we tracked 300000 macroparticles interacting through the estimated machine wakefield [1] over 4 damping times. Such a high number of the macropatricles has been proven to be necessary to avoid artificial numerical growth of the bunch energy spread.

Finally, the bunch length has been measured in the DAΦNE positron ring. The bunch signal was picked up by a broad band button [2]. The resulting bunch distribution has been found by processing the button signal taking into account the button transfer impedance and the attenuation of the cables connecting the button to a sampling oscilloscope.

Below we describe the results of the study and compare the analytical predictions with numerical and measured data. The main relevant DAΦNE main ring parameters are listed in Table 1.

**Table 1: Main DAΦNE Parameters**

| | | | |
|---|---|---|---|
| Energy | E | 510.0 | Mev |
| Average radius | R | 15.548 | m |
| Emittance | $\varepsilon_x$ / $\varepsilon_y$ | 1/0.01 | mm·mrad |
| Beam-beam tune shift | $\xi_x$ / $\xi_y$ | 0.04/0.04 | |
| Betatron tune | $\nu_x$ / $\nu_y$ | 5.09/5.07 | |
| RF frequency | $f_{rf}$ | 368.26 | MHz |
| Harmonic number | h | 120 | |
| Revolution frequency | $f_0$ | 3.0688 | MHz |
| Max. number of bunches | $n_b$ | 120 | |
| Minimum bunch separation | $s_b$ | 81.4 | cm |
| Bunch average current | $I_0$ | 43.7 | mA |
| Particles per bunch | N | $9.0 \cdot 10^{10}$ | |
| Momentum compaction | $\alpha_c$ | 0.017 | |
| Natural energy spread | $\sigma_{\varepsilon 0}$ / $E$ | 0.000396 | |
| Bunch length | $\sigma_z$ | 2.0 - 3.0 | cm |
| Synchrotron radiation loss | $U_0$ | 9.3 | keV/turn |
| Damping time | $\tau_\varepsilon$ / $\tau_x$ | 17.8/36.0 | ms |
| RF voltage | $V_{rf}$ | 100 - 250 | kV |
| Synchrotron tune | $\nu_s$ | 0.011 | |
| Beta functions at IP | $\beta_x^*$ / $\beta_y^*$ | 450/4.5 | cm |
| Maximum luminosity | L | $5.3 \cdot 10^{32}$ | $cm^{-2}s^{-1}$ |

## 2. Analytical estimates

It is widely believed (see [3, 4], for example) that the microwave instability is caused by the bunch longitudinal coherent mode coupling. The instability can manifest itself either through the coupling among the azimuthal modes or the radial ones having the same azimuthal number.

In the first case the frequencies of the azimuthal modes have to be shifted by amounts comparable to the synchrotron frequency (“strong” instability). The frequency shifts leading to the radial mode coupling can be substantially smaller than the synchrotron frequency (“weak” instability).

An analytical model, which we follow in this paper, allows to treat the mode coupling taking into account splitting of each azimuthal mode in two radial modes. It is based on the approximation of the real bunch distribution by a double Water Bag distribution [5].

In the angle-action phase space such a distribution is described by the equation:

$$\psi(J) = \overline{\psi}\left[(1-\Gamma)U\left(J_1 - J\right) + \Gamma U\left(J_2 - J\right)\right] \tag{1}$$

where the constant $\overline{\psi}$ is derived from the normalization condition, $\Gamma$ is a parameter which has to be chosen between 0 and 1 to better approximate the real distribution by the double Water Bag, $U(J)$ is the step function, and $J$ the action variable.

By substititing the above distribution in the Vlasov equation, the following eigenvalue system is obtained [see Appendix for details]:

$$\begin{cases} \left[\Omega - m\omega\left(J_1\right)\right]A_m = -(1-\Gamma)\displaystyle\sum_{l=-\infty}^{\infty}\left(M_{11}^{ml}A_l + M_{12}^{ml}B_l\right) \\ \left[\Omega - m\omega\left(J_2\right)\right]B_m = -\Gamma\displaystyle\sum_{l=-\infty}^{\infty}\left(M_{21}^{ml}A_l + M_{22}^{ml}B_l\right) \end{cases} \tag{2}$$

with

$$M_{ij}^{ml} = i\frac{\alpha_c e^2 N}{8\pi^3 T_o E\left[(1-\Gamma)J_1 + \Gamma J_2\right]\omega\left(J_i\right)} \int_0^{2\pi}\varepsilon\left(J_i,\phi\right)e^{-im\phi}d\phi\int_0^{2\pi}e^{il\phi'}w\left[z\left(J_i,\phi\right) - z\left(J_j,\phi'\right)\right]d\phi' \tag{3}$$

The frequencies of the coherent modes $\Omega_{m,k}$, with $m$ being the azimuthal mode number and $k$ the radial one ($k$ = 0;1), are obtained by equating to 1 the determinant of the eigenvalue system. The imaginary part of $\Omega m,k$ gives the rise time of the weak microwave instability. Such an imaginary part comes out when two radial modes couple together. The quantity $\omega(J)$ is the synchrotron frequency depending on $J$ due to the non linearities of the wake fields $w(z)$. The other quantities are described in Appendix together with the derivation of the system (2) from the Vlasov equation. Actually, the eigenvalue system has an infinite number of equations ($m$ = 0,…, ∞) which has to be truncated to get $\Omega_{m,k}$. In our case we limit our analysis up to $m$ = 9.

The machine wake function $w(z)$ we used in the analytical model was calculated at the project stage of the collider [1]. Figure 1 shows the wake potential of a short Gaussian bunch with the rms length $\sigma_z$ = 2.5 mm, i. e. much shorter than the nominal bunch length of 3 cm, which we use as the machine wake function.

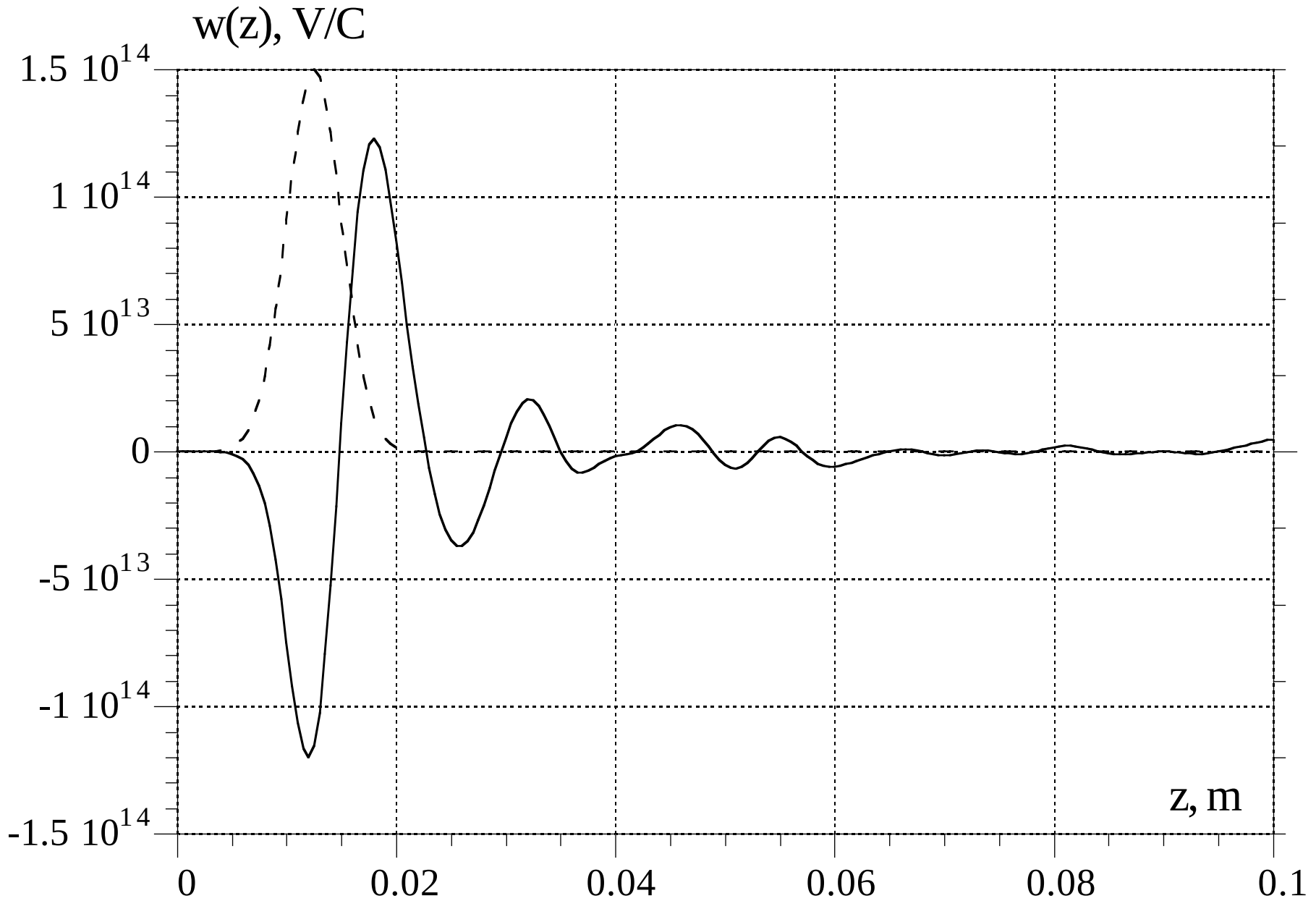

Figure 1: DAΦNE wake potential of a 2.5 mm Gaussian bunch.

The first nine azimuthal modes each splitting in two radial lines are shown in Fig. 2. Here the dependencies of the mode frequencies on the bunch current are given for four possible RF voltages (100 kV, 150 kV, 200 kV and 250 kV). The respective decrements (inverse rise times) of the instability due to the radial mode coupling are reported in Fig. 3.

As it can be seen, at lower voltages the instability threshold is caused by the coupling of the modes with low azimuthal numbers. For example, the instability at $V_{RF}$ = 100 kV is driven by the radial mode coupling of the quadrupole (m = 2) and sextupole modes (m = 3). At higher voltages the coupling of the higher order modes results in an instability. Namely, at 200 kV and 250 kV the first unstable modes are m = 9 and m = 8, respectively.

The unstable longitudinal coherent modes cause longitudinal bunch shape modulation. This is a harmful effect because the modulation excites new beam-beam resonances degrading the collider performance. Therefore, higher RF voltages would be preferable for DAΦNE operation since the higher order unstable modes create higher order beam-beam resonances which are much less dangerous for the beam-beam interaction.

Figure 4 enlarges the scale of Fig. 2 to show in details the coupling of the first longitudinal modes (dipole, quadrupole and sextupole) which are supposed to be most dangerous for the beam-beam collisions. At 100 kV the mode defining the microwave instability threshold at the current of 25 mA is the quadrupole one ($m$ = 2). By increasing the RF voltage to 150 kV the threshold is pushed up to 38 mA.

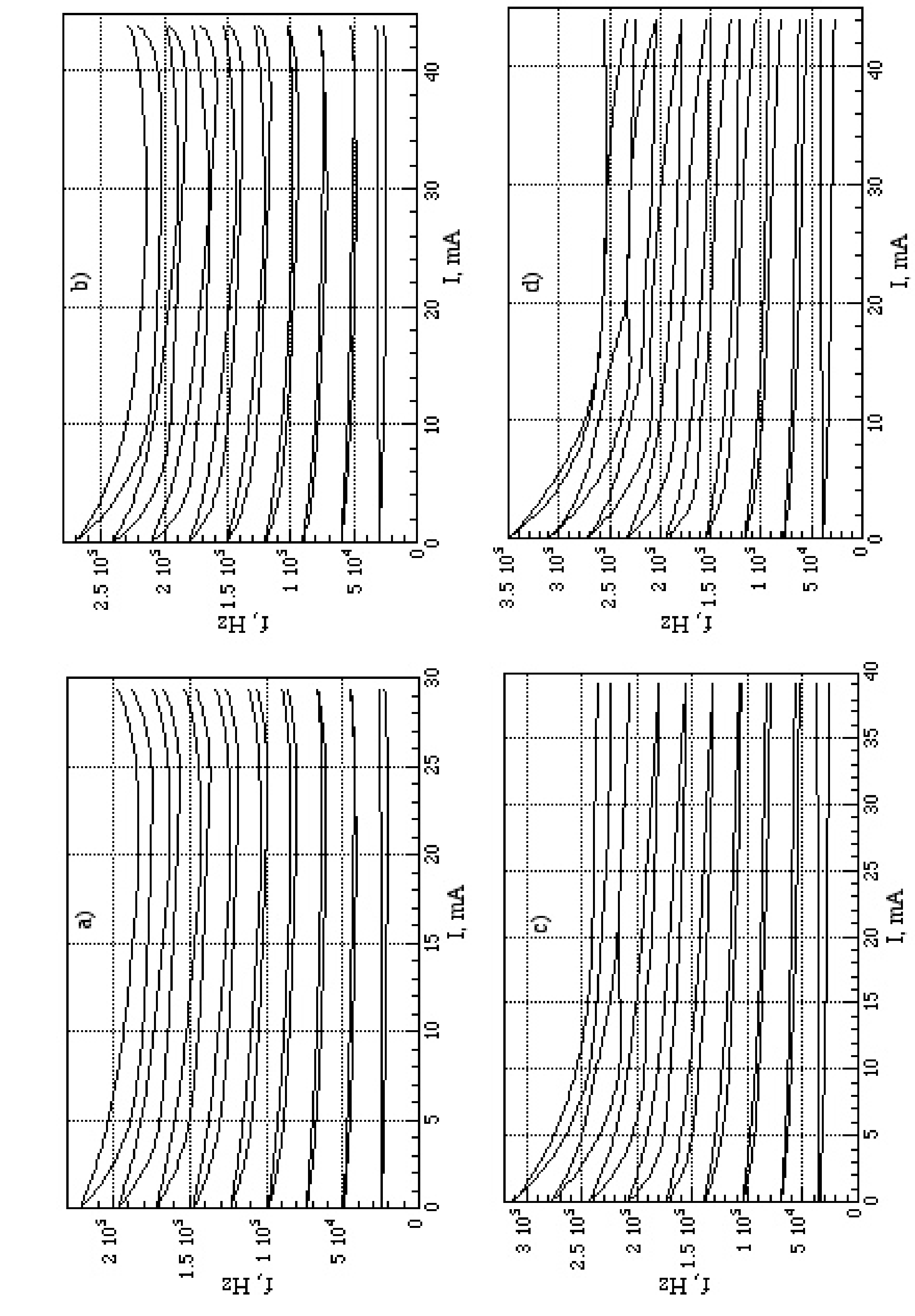

Figure 2: Longitudinal coherent mode coupling at different RF voltages - a) 100 kV; b) 150 kV; c) 200 kV; d) 250 kV

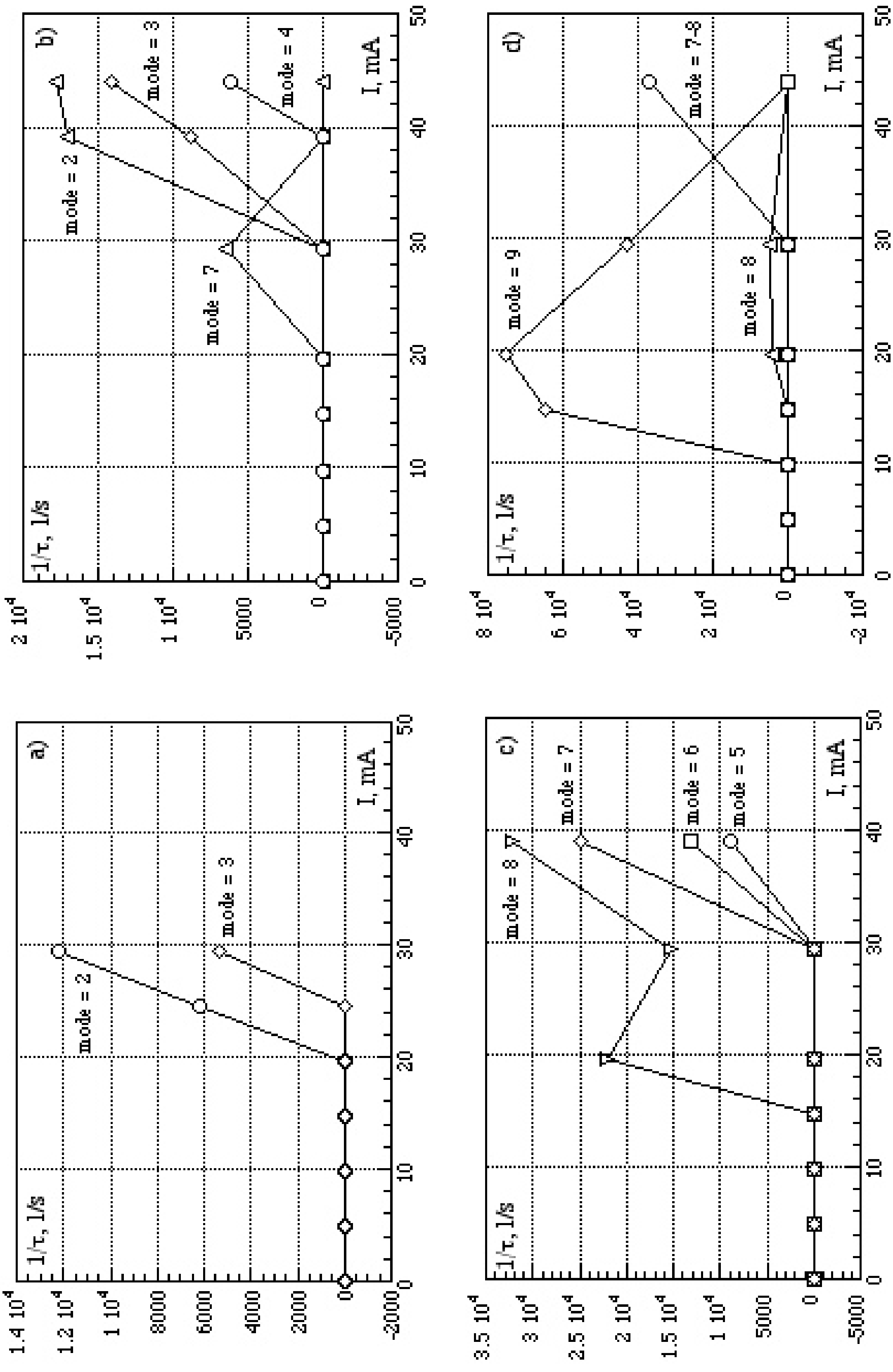

Figure 3: Microwave instability decrements (inverse rise times) at different RF voltages - a) 100 kV; b) 150 kV; c) 200 kV; d) 250 kV

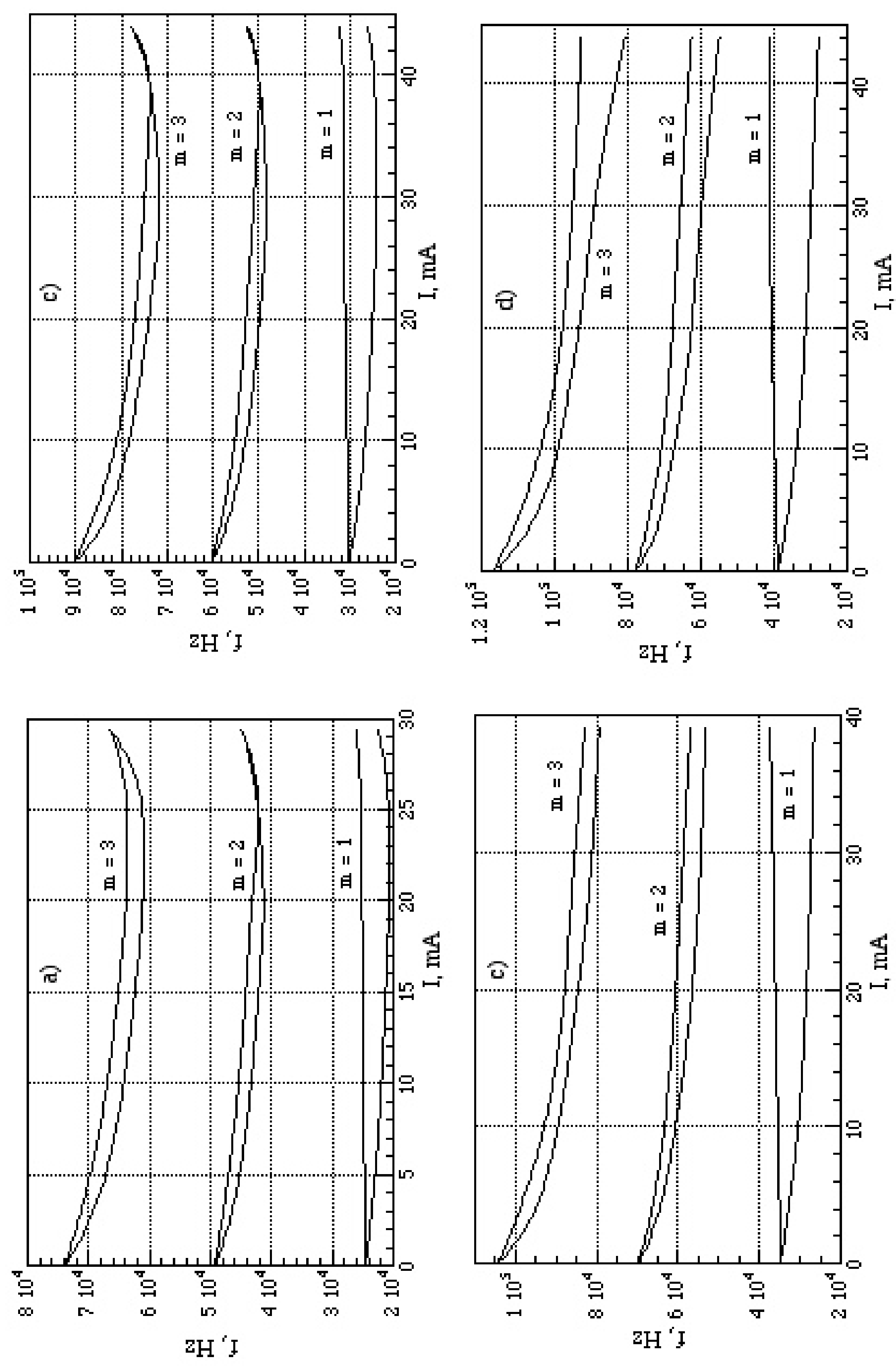

Figure 4: Dipole (m = 1), quadrupole ( m = 2) and sextupole (m = 3) mode coupling at different RF voltages.
a) 100 kV; b) 150 kV; c) 200 kV; d) 250 kV

This coincides exactly with our observations during the bunch lengthening measurements. We could detect an appearance of the pure quadrupole synchrotron sidebands at 25 - 26 mA with $V_{RF}$ = 100 kV. The dipole mode got unstable later at about 35 mA. An RF voltage increase to 150 kV shifted the quadrupole mode threshold to 38 mA, while the dipole mode was stable up to the nominal bunch current of 44 mA. For higher RF voltages the coupling of the low coherent modes should occur for currents higher than the nominal value.

However, we should note here that the instability growth time is much faster than the machine natural damping time of 17 ms for both the lower and higher order coherent modes. This means that the instability threshold is lower for higher RF voltages even though the instability is excited by the coupling of higher order unstable modes. This has been checked by numerical simulations.

## 3. Numerical simulations

In order to simulate the bunch lengthening process, we have undertaken a numerical tracking using the wake potential of a short gaussian bunch with $\sigma_z$ = 2.5 mm shown in Fig. 1 as the machine wake function.

The tracking method is essentially the same as that successfully used in the bunch lengthening simulations for the SLC damping rings [6, 7], SPEAR [8], PETRA and LEP[9] and DAΦNE accumulator ring [10]. The motion of $N_s$ superparticles representing the beam is described in the longitudinal phase space by [7]:

$$\varepsilon_i(n) = \varepsilon_i(n-1) - \frac{2T_0}{\tau_\varepsilon}\varepsilon_i(n-1) + 2\sigma_{\varepsilon 0}\sqrt{\frac{T_0}{\tau_\varepsilon}}R_i(n) - U_0 + \hat{V}\cos\left[\phi_s - \frac{2\pi h}{L_0}z_i(n)\right] + V_{ind}[z_i(n)] \quad (4)$$

$$z_i(n) = z_i(n-1) + \frac{\alpha c T_0}{E}\varepsilon_i(n)$$

where $\varepsilon_i(n)$ and $z_i(n)$ are the energy and position coordinates of the $i^{\text{th}}$ particle after n revolutions in the storage ring. $T_0$ is the revolution period; $\tau_\varepsilon$ the damping time; $U_0$ the energy lost per turn; $\phi_s$ the synchronous phase; $h$ the harmonic number; $L_0$ the machine length; $R_i$ a random number obtained from a normally distributed set with mean 0 and rms 1.

On each turn all the super particles are distributed in $N_{bin}$ bins and the induced voltage $V_{ind}$ is calculated by [7]:

$$V_{ind}(z_j) = -\frac{Q}{N_s}\sum_{z_i<z_j}^{i=1,N_{bin}} N_b(z_i)w(z_j - z_i) \quad (5)$$

Note that $z_j$ in the expression (5) are the coordinates of the bin centers and the induced voltage at the positions of the super particles is found by a linear interpolation between the $V_{ind}(z_j)$. Here $N_b(z_i)$ is the number of super particles in the bin with the center at $z_i$ and $w(z)$ is the machine wake function.

In our simulations 300000 macroparticles particles are tracked over 4 damping times and the average bunch properties, as rms length, rms energy spread, coordinate of the centroid, are calculated by averaging over the last damping time. Such a high number of macroparticles is necessary to avoid artificial numerical bunch energy spread growth.

Figure 5 shows an example of the numerical energy spread growth as a function of modelling macroparticles for the bunch current of 30 mA. 300000 particles have been chosen as a compromise between an acceptable accuracy and a reasonable CPU time (still it takes some days of CPU time on the LNF UNIX computer cluster to perform the simulations). If has been also found that the results do not change much if the number of bins is higher than 80.

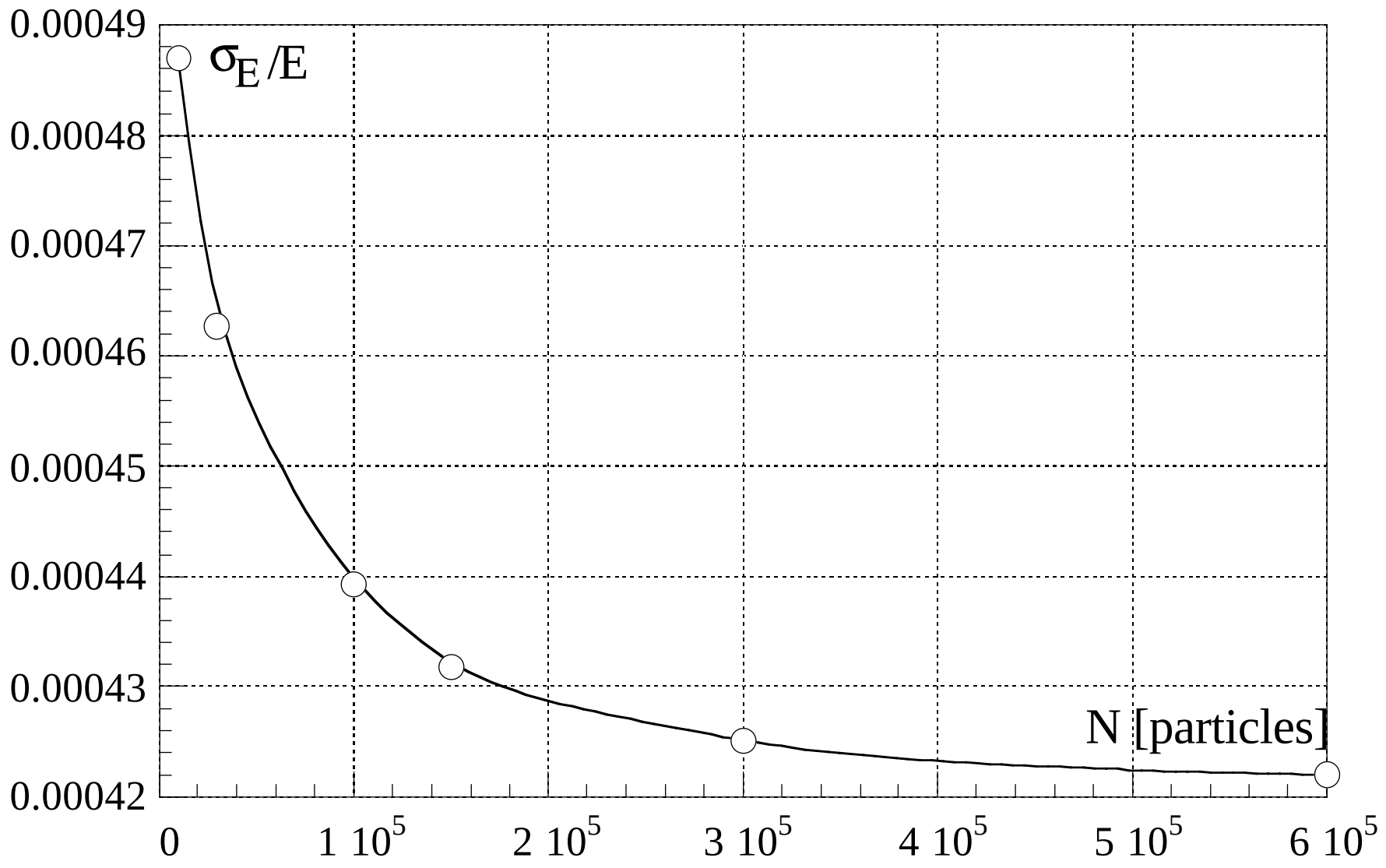

Figure 5: Numerical energy spread growth as a function of number of superparticles.

The results of the numerical simulations for the DAΦNE main rings are summarised in Fig. 6. In particular, this Figure shows the rms bunch length (a), rms bunch energy spread (c), the bunch centroid shift (b) as a function of the bunch current, calculated for two RF voltages, 100 kV and 250 kV, which are considered as limits of the RF cavity working range.

The normalised bunch distributions are shown in the last plot (d) in Fig. 6. The distributions are wider than Gaussian distributions due to the bunch interaction with the inductive machine impedance and slightly distorted due to the real impedance. Because of this, we will use the full width at half maximum size (FWHM) of the bunch when comparing the simulation results and measurement data.

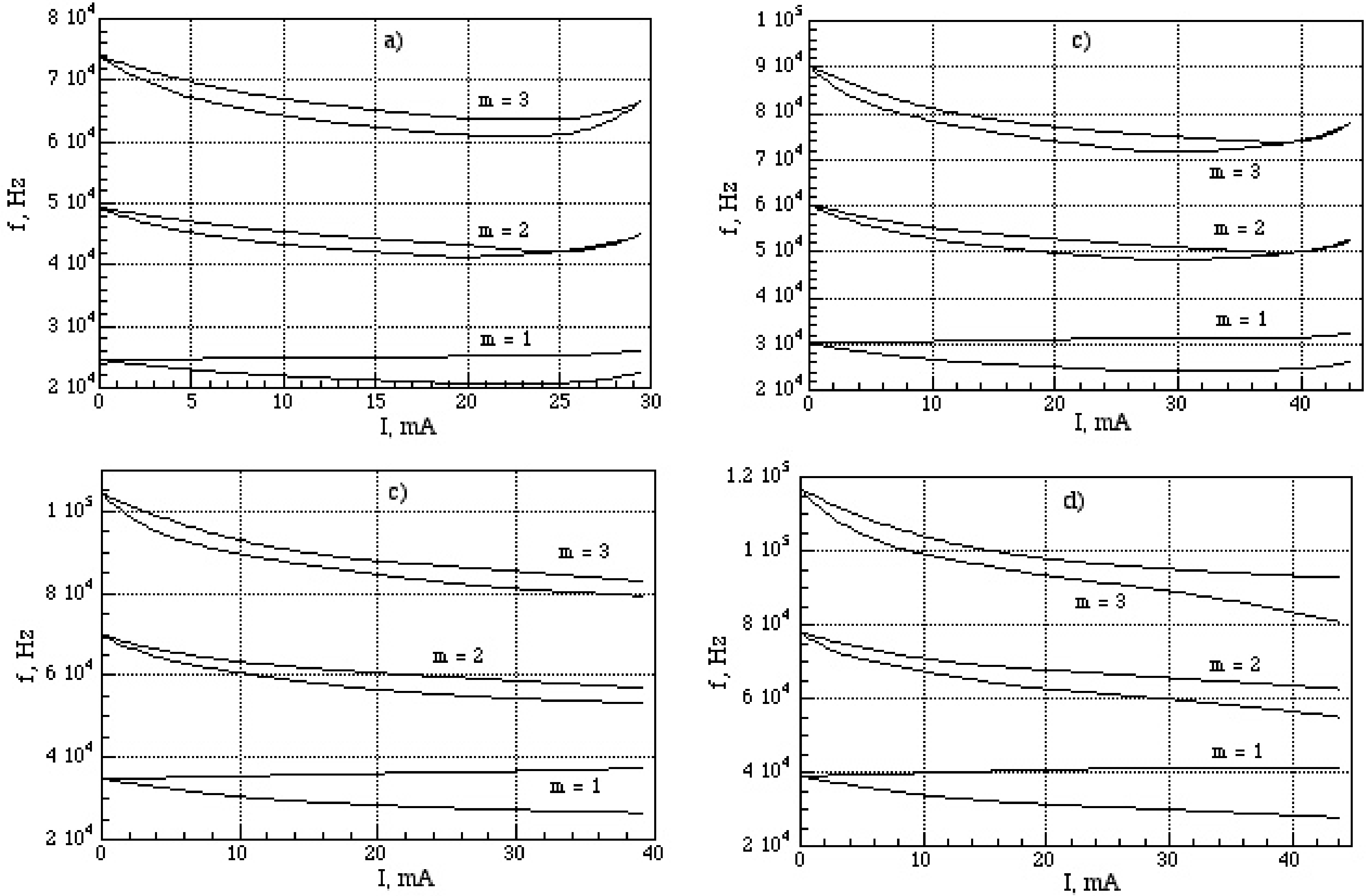

Figure 6: Results of numerical simulations at RF voltage of 100 kV and 250 kV - a) rms bunch length; b) bunch centroid; c) rms energy spread; d) bunch distribution at nominal current.

As expected (see the discussion in the previous chapter), the microwave instability threshold (see Fig. 6(c)) is lower and the bunch energy spread is higher for the higher RF voltage. In both cases the bunch widening is modest: the bunch energy spread growth does not exceed 40% of the nominal value.

As it is seen the numerical thresholds are by about 30% lower than the calculated ones. In our opinion, this is still a good result considering that the analytical model is rather crude. One could refine the model including splitting of each azimuthal mode in more radial lines. The agreement could probably be improved if the real machine wake function is used in the model instead of the wake potential of a very short, but still finite , Gaussian bunch. But this is almost an impossible task since no one numerical code calculates this function directly.

## 4. Bunch length measurements

The bunch length in the DAΦNE positron ring has been measured processing the beam signal from a broad band button electrode [2] connected with a low attenuation cable (ANDREW FSJB-50A) 8 m long to a sampling oscilloscope TEKTRONIX 11801A, equipped with a sampling head SD-24 with a rise time of 17.5 psec and an equivalent bandwidth of 20 GHz. The stability of the waveform has been achieved by using the signal from a stripline as a trigger. The waveform is sent via a GPIB interface to the control system for storage and off - line reconstruction.

The resulting bunch distribution is found by processing the button signal taking into account the button transfer impedance and attenuation of the cable connecting the button to the measuring device.

Let $I(\omega)$ and $V(\omega)$ be the Fourier transform of the real bunch distribution and the signal recorded by the oscilloscope, respectively. The two values are related by:

$$V(\omega) = I(\omega) Z_b(\omega) \alpha(\omega) \quad (6)$$

where $Z_b(\omega)$ is the button transfer impedance and $\alpha(\omega)$ is the frequency dependent cable attenuation. Then, the bunch distribution in time domain is obtained by the inverse Fourier transform of $I(\omega)$ as:

$$I(t) = \frac{1}{2\pi} \int_{-\infty}^{+\infty} \frac{V(\omega)}{Z_b(\omega)\alpha(\omega)} e^{j\omega t} d\omega \quad (7)$$

### *4.1 Button transfer function*

It has been shown that the DAΦNE broad band button electrode transfer impedance is described well up to the frequency of 5.2 GHz by the following analytical formula [2]:

$$Z_b(\omega) = \phi R_0 \left( \frac{\omega_1}{\omega_2} \right) \frac{j\omega / \omega_1}{1 + j\omega / \omega_1} \tag{8}$$

where $\omega_1 = 1/(R_0 C_b)$, $\omega_2 = c/2r$ and the coverage factor $\phi = r/4b$. In our case the capacitance $C_b$ of the button to ground is equal to 3.8 pF, $R_0$ is 50 Ω, the button radius $r$ is 7.5 mm and the radius of the beam pipe $b$ is equal to 44 mm.

Beyond 5 GHz the wave lengths get comparable with the button size and the button transfer impedance is no longer described well by the analytical expression (2). Indeed, at 5.2 GHz we observe the first button resonance which corresponds to the TE-111 mode trapped around the button. Further numerical simulations have also shown that at higher frequencies the button transfer impedance gets lower and tends to zero at approximately 7 GHz. This means that we can rely on the measurements results only for the bunches whose frequency spectrum lies within 5 GHz, i. e. for the bunches longer than ~ 1.5 cm. Because of that, in the following we will compare the numerical simulations and measurement results obtained at relatively low RF voltages.

Figure 7a) shows the button response to a Gaussian pulse which we can get applying eq. (8), while Fig. 7b) shows a typical measured signal.

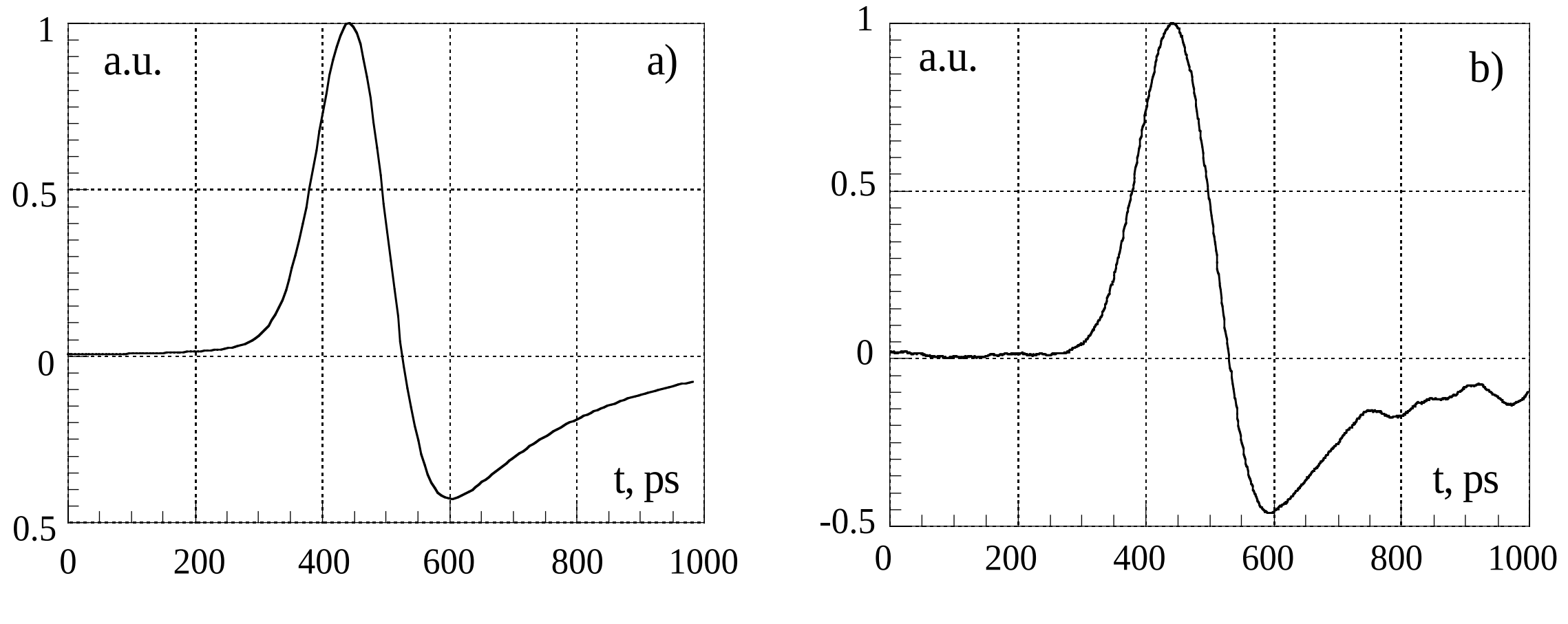

Figure 7: Button response: a) Theoretical button response to a Gaussian pulse;
b) Example of a measured signal with the button.

As it can be seen, the shape of the two signals are very similar, except for a small ringing at the tail of the measured signal. It corresponds to the oscillations with the frequency of 5.2 GHz, i. e. the tail of the bunch spectrum couples to the first button resonance at this frequency. However, this does not affect much the results of our measurements, especially for longer bunches.

### *4.2 Cable attenuation*

In order to reduce the signal distortion as much as possible one has to choose a cable with appropriate parameters: low attenuation, high maximum cable frequency, i. e. cut-off frequency of the first parasitic propagating wave, and minimal cable length.

It turns out that due to the space constraints the minimal allowable cable length is 8 m for the bunch length measurements in the DAΦNE positron ring, while it is equal to 13 m for the electron ring.

We have choosen a low attenuation 1/4” ANDREW FSJB - 50 A cable. We could also select a 1/2’’ ANDREW FSJB - 50 B which has even lower attenuation, but it has lower maximum frequency (10 GHz) and, what is more important, having bigger diameter, it is more difficult to pull it between the control room and the BPM.

Below we estimate the possible signal distortion due to the cables, i.e. the signal elongation and its shape distortion.

The cable Green function, i. e. the cable response to the δ-signal is given by [11]:

$$h(t) = A\,\boldsymbol{exp}\left(-\frac{B}{u(t)}\right)[u(t)]^{-3/2} \tag{9}$$

where

$$u(t) = \begin{cases} t - l/\beta c; & u(t) \geq 0 \\ 0; & u(t) < 0 \cdot \end{cases} \tag{10}$$

Here $l$ is the cable length and $\beta c$ is the velocity of the signal propagation along the cable:

$$\beta c = 1/\sqrt{L'C'} \tag{11}$$

where *L’* and *C’* are the cable inductance and capacitance per meter, respectively. The coefficients *A* and *B* are given by:

$$A = \sqrt{B/\pi} \tag{12}$$

$$B = \frac{1}{\pi f}\left(\frac{\alpha[dB/m]l}{17.39}\right)^2 \tag{13}$$

with $f$ the frequency and $\alpha$ the cable attenuation. Note, since $\alpha$ is measured in *dB/m* , it scales as ~ $f$1/2 (resistive wall losses) and *B* does not depend on the frequency.

The signal at the cable exit *S(t)* can be found by a convolution of the Green function over the time dependent signal at the cable entrance *s(t)* as:

$$S(t) = \int_0^t s(t')h(t-t')dt' \tag{14}$$

In particular, the cable response to a step signal is:

$$S(t) = 1 - \Phi\left(\sqrt{\frac{B}{u(t)}}\right) \tag{15}$$

where Φ(x) is the error function. The expression (15) allows to estimate how a cable rise time depends on the cable length and attenuation. As an example, in Fig. 8 we compare responses to a step signal of the 1/4'' FSJ1-50A cables of 8 m (a) and 60 m (b) long.

One can also easily find a cable response to a rectangular pulse of a duration $\Delta t$ as the difference of the cable responses to the two step signals separated by the time interval $\Delta t$.

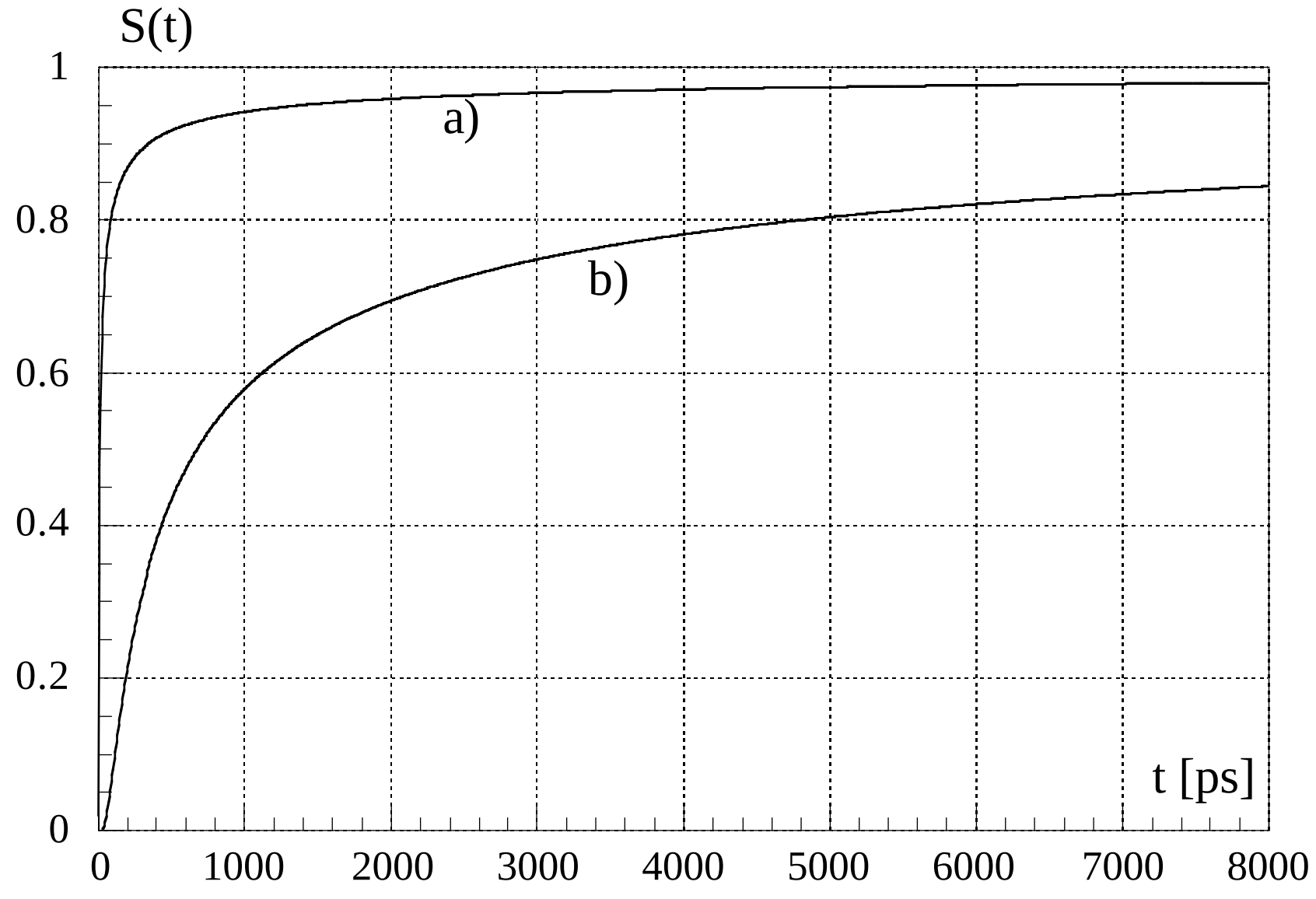

Figure 8: 1/4'' FSJ1-50A cable response to a step signal:
a) 8 m long cable; b) 60 m long cable.

However, the bunch shape is expected to be nearly Gaussian in the DAΦNE main rings. In order to evaluate how strong the distortion due to the cables could be, let us consider a cable response to a purely Gaussian pulse with rms length σ. Then, the measured signal is proportional to:

$$S(\tau) \sim \int_0^{\tau} \boldsymbol{exp}\left\{-\frac{1}{2}(\tau'-5)^2\right\} \boldsymbol{exp}\left\{-\frac{Bc}{\sigma}\frac{1}{(\tau-\tau')}\right\}(\tau-\tau')^{-3/2} d\tau' \tag{16}$$

Here we use the normalized variables $\tau = t/\sigma$ and $\tau' = t'/\sigma$.

Figure 9 shows the responses to the Gaussian pulse of 100 ps (bunch length $\sigma_Z$ = 3 cm) for cables 8 m (b) and 60 m (a) long, respectively. As it can be seen, the measured signal for the 60 m long cable is strongly distorted and much wider with respect to the actual Gaussian one, while the 8 m long cable signal closely resembles the primary Gaussian pulse.

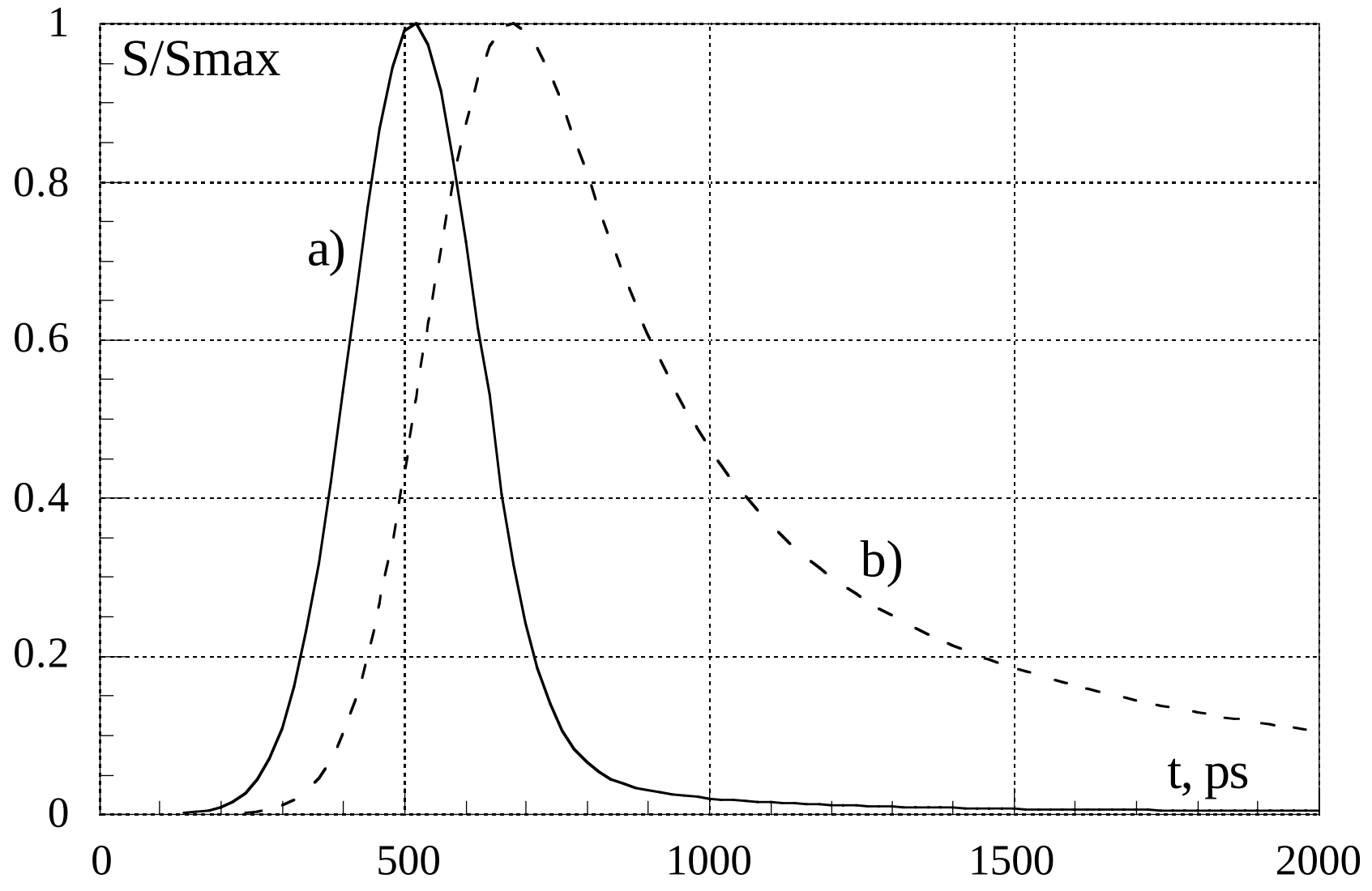

Figure 9: 1/4'' FSJ1-50A cable response to a 100 ps Gaussian signal:
a) 8 m long cable; b) 60 m long cable.

In order to give a quantitative measure of the signal elongation due to the cable attenuation let us introduce a "distortion factor" *F* as a ratio of the rms value of an input Gaussian pulse to the full width at half maximum value (FWHM) of the output pulse normalized to $2\sqrt{(2\ln 2)}$ = 2.3548... Clearly, in the absence of cable attenuation (distortion) the factor *F* is equal to unity.

For the example of Fig. 9, the distortion factor is equal to 1.064 and 1.96 for the 8 m long cable and 60 m long cable, respectively, i. e. for the longer cable the measured signal is longer by about a factor of 2 than the Gaussian pulse at the cable input.

Table 2 summarizes the distortion factors for two cables, 1/4'' FSJ1-50A and 1/2'' FSJ4-50B, calculated for the expected range of the bunch length in the main rings ($\sigma_Z$ = 1 - 3 cm).

**Table 2: Distortion factor due to cables in bunch length measurements**

| | FSJ1 - 50A cable | | FSJ4 - 50B cable | |
|---|---|---|---|---|
| $\sigma_Z$ , cm | electron ring | positron ring | electron ring | positron ring |
| 1 | 1.2286 | 1.1375 | 1.1375 | 1.0910 |
| 2 | 1.1547 | 1.0976 | 1.0976 | 1.0510 |
| 3 | 1.1198 | 1.0643 | 1.0643 | 1.0422 |

As it can be seen, the distortion introduced by both cables does not differ much. So, we have chosen FSJ1 - 50A cable since it is more mechanically flexible and we have calibrated the 8 m long piece of the cable measuring the dependence of the attenuation on the frequency (see Fig. 10).

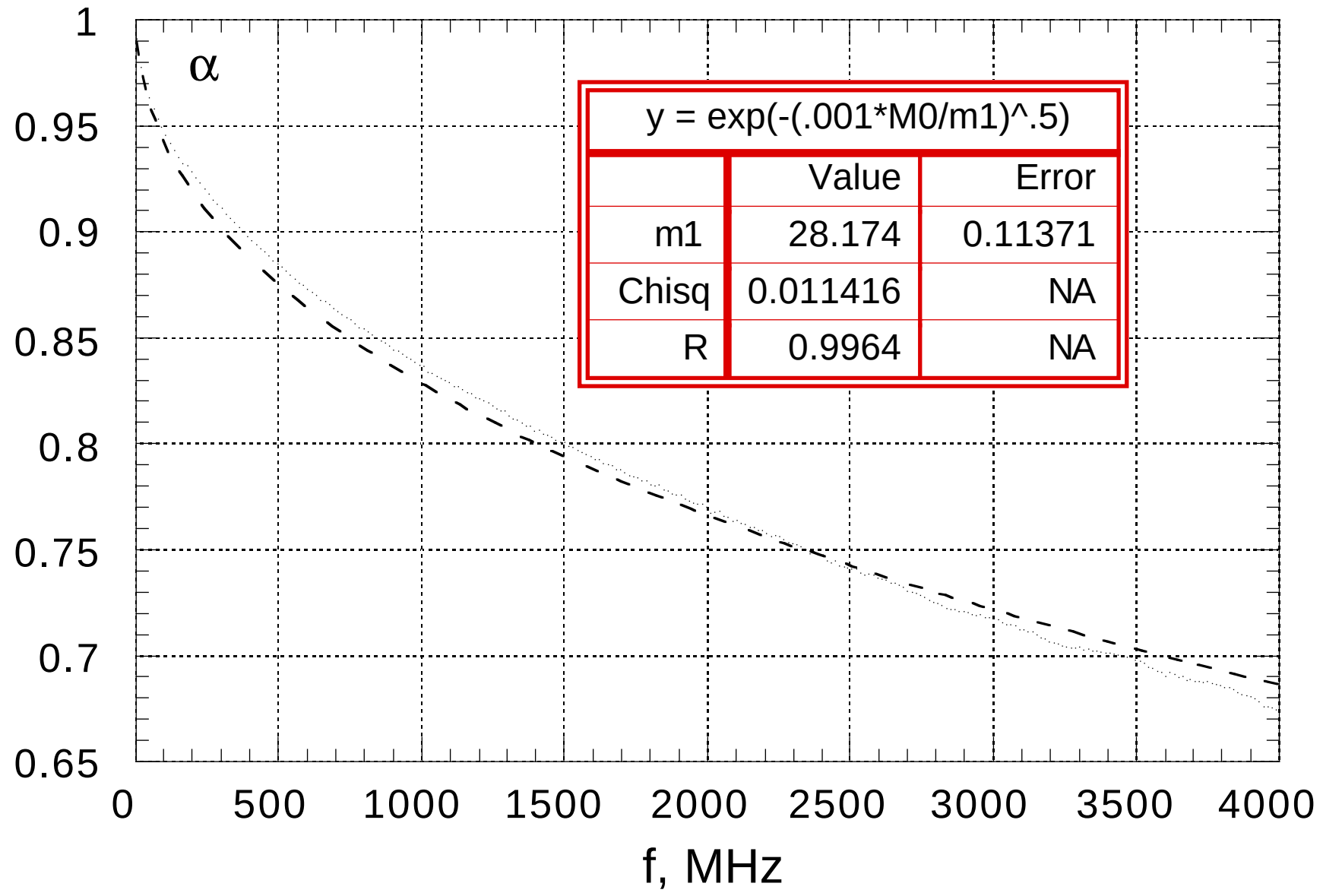

Figure 10: Dependence of the cable attenuation on frequency.
Solid line - measurements; dotted line - analytical fit.

### *4.3 Measurement results*

First, we recorded bunch signals at low RF voltages of 100 kV and 150 kV because we were expecting that the bunch would be long enough to have the bunch spectrum below 5 GHz. In this frequency range the button transfer impedance is described well by the analytical formula (see discussion in 4.1). In order to elaborate the bunch distributions from the measured data, we apply eqs. (6) - (7) taking into account the button transfer impedance, eq.(8) and the frequency dependent cable attenuation shown in Fig. 10.

The measurements were performed for the currents in the range from ~ 0 mA to 48 mA (note that the nominal design current is 43.7 mA). Figure 11 shows an example of the elaborated bunch distributions in case of $V_{RF}$ = 150 kV for the currents of 2 mA (a) and 48 mA (b), respectively.

It is clearly seen that the bunch gets wider at higher current due to the interaction with the imaginary (mainly inductive) machine impedance and the distribution symmetry is slightly broken because of power losses from the real part of the coupling impedance.

For shorter bunches (Fig. 11(a)) we observe oscillations in the tail of the distribution which correspond to the first parasitic button resonance. This means that at low currents the bunch is still relatively short such that the tail of its frequency spectrum couples to this resonance. For longer bunches (see Fig. 11(b)) the oscillations disappear.

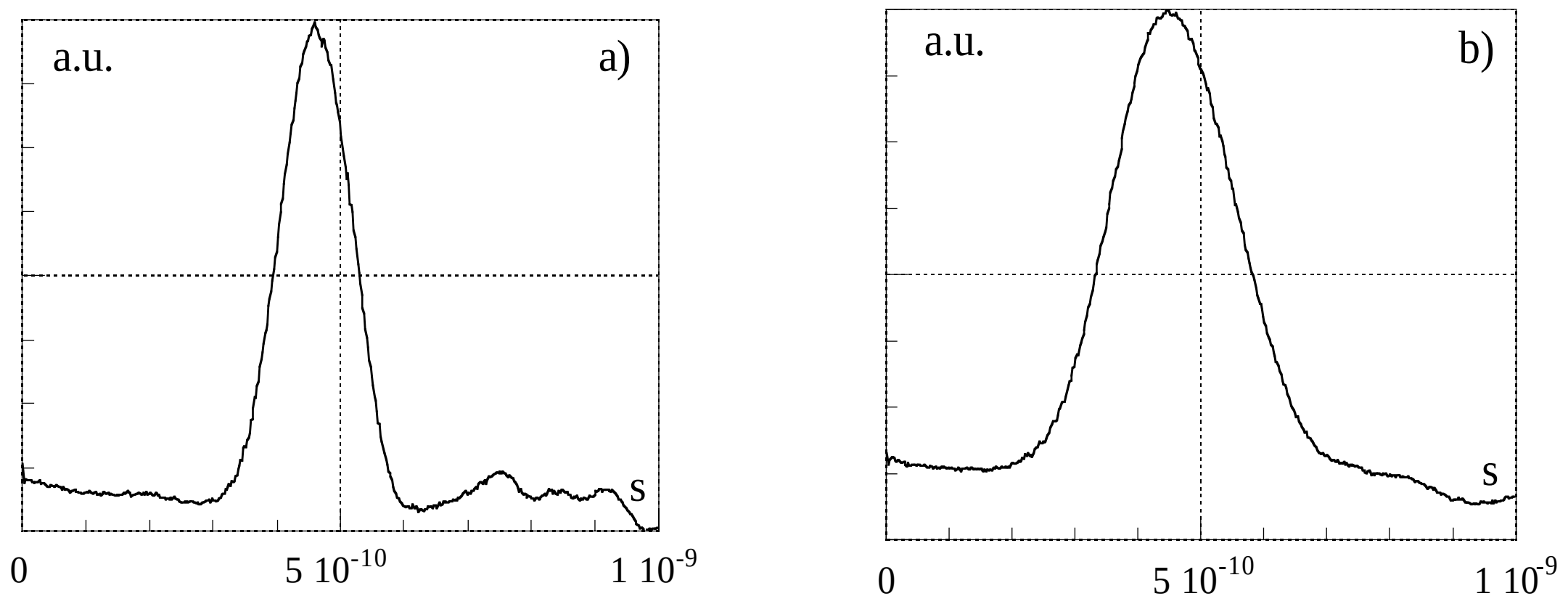

Figure 11: Bunch distribution at 150 kV: a) bunch current 2 mA; b) bunch current 48 mA.

Figure 12 compares the measured bunch length (FWHM, circles) with the results of numerical simulations (solid line). The agreement is very much satisfactory in almost all range of currents, except for low currents < 5 mA.

Certainly, the discrepancy at the low currents is not due to errors of the numerical simulations since the natural bunch length at the given voltage is known quite precisely and agrees well with the simulations. We attribute the small disagreement to the above discussed fact that for frequencies higher 5. 2 GHz the button transfer impedance is no longer described well by the analytical expression (8) and, therefore, our treatment is not valid for bunches shorter than ~1.5 - 1.7 cm.

However, we believe that it is safe to rely on the results of the numerical simulations in this case. Moreover, the simulations reproduce well not only the bunch length behavior as a function of current, but also the internal bunch distribution. Figure 13 shows as an example a comparison between the measured and simulated bunch shapes at I = 26 mA and $V_{RF}$ = 100 kV. This means that the estimated wake field, shown in Fig. 1, can be successfully applied for the bunch length (and shape) evaluations at different working conditions (bunch currents, RF voltages, momentum compactions etc.).

Fourier transform of the wake wake fields demonstrates that in the frequency range up to 20 GHz the absolute value of the normalized coupling machine impedance $|Z/n|$ does not exceed 0.6 Ω.

As a final result, we plot the bunch lengthening graphs (Fig. 14) calculated numerically at two limiting RF voltages of 100 kV and 250 kV to show a possible range of bunch length variations. Here we show FWHM bunch length instead of the rms size since the bunch shape differs from Gaussian.

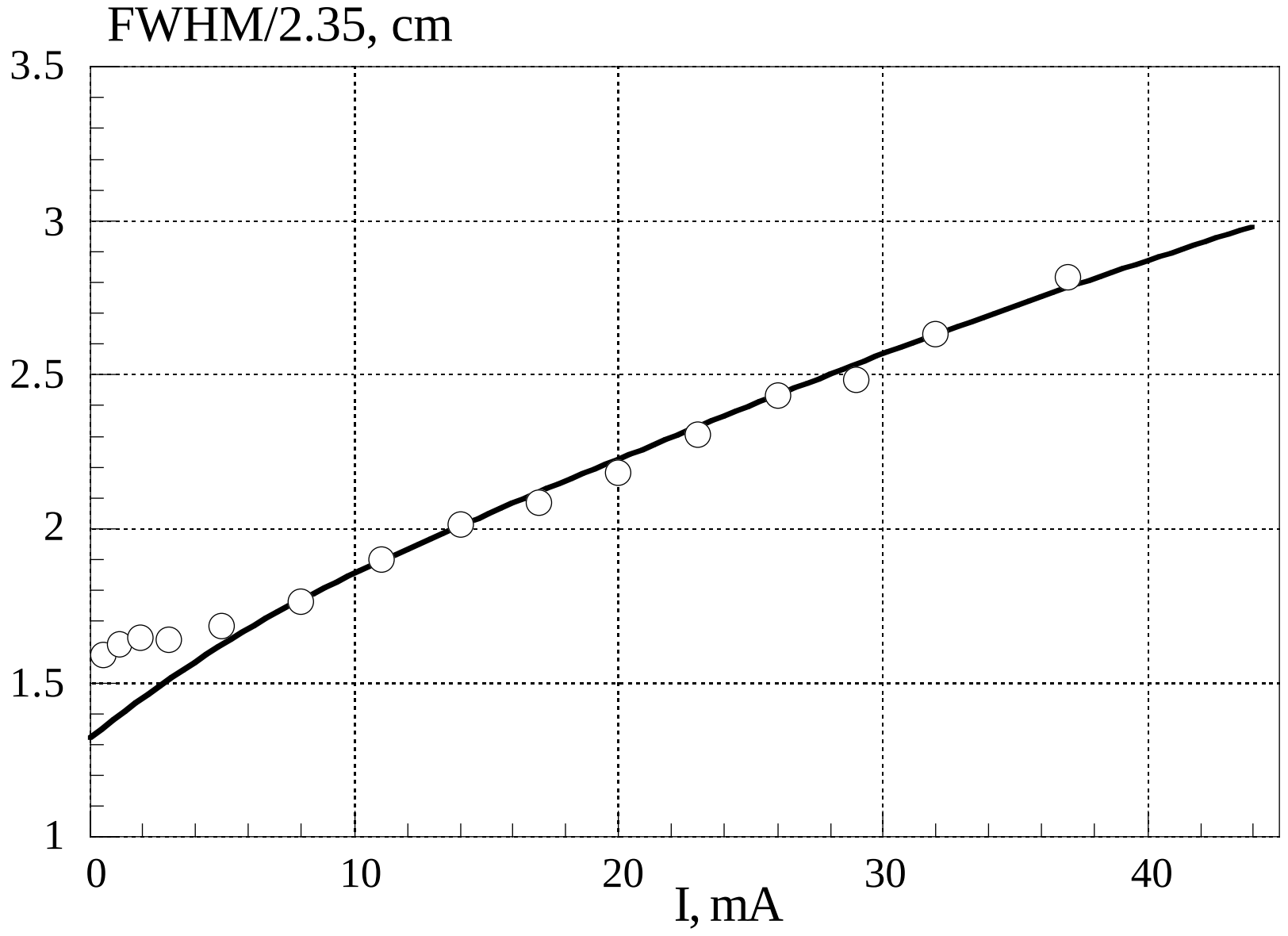

Figure 12: Bunch lengthening (FWHM) at 100 kV RF voltage.
Solid line - numerical calculations; circles - measurement results.

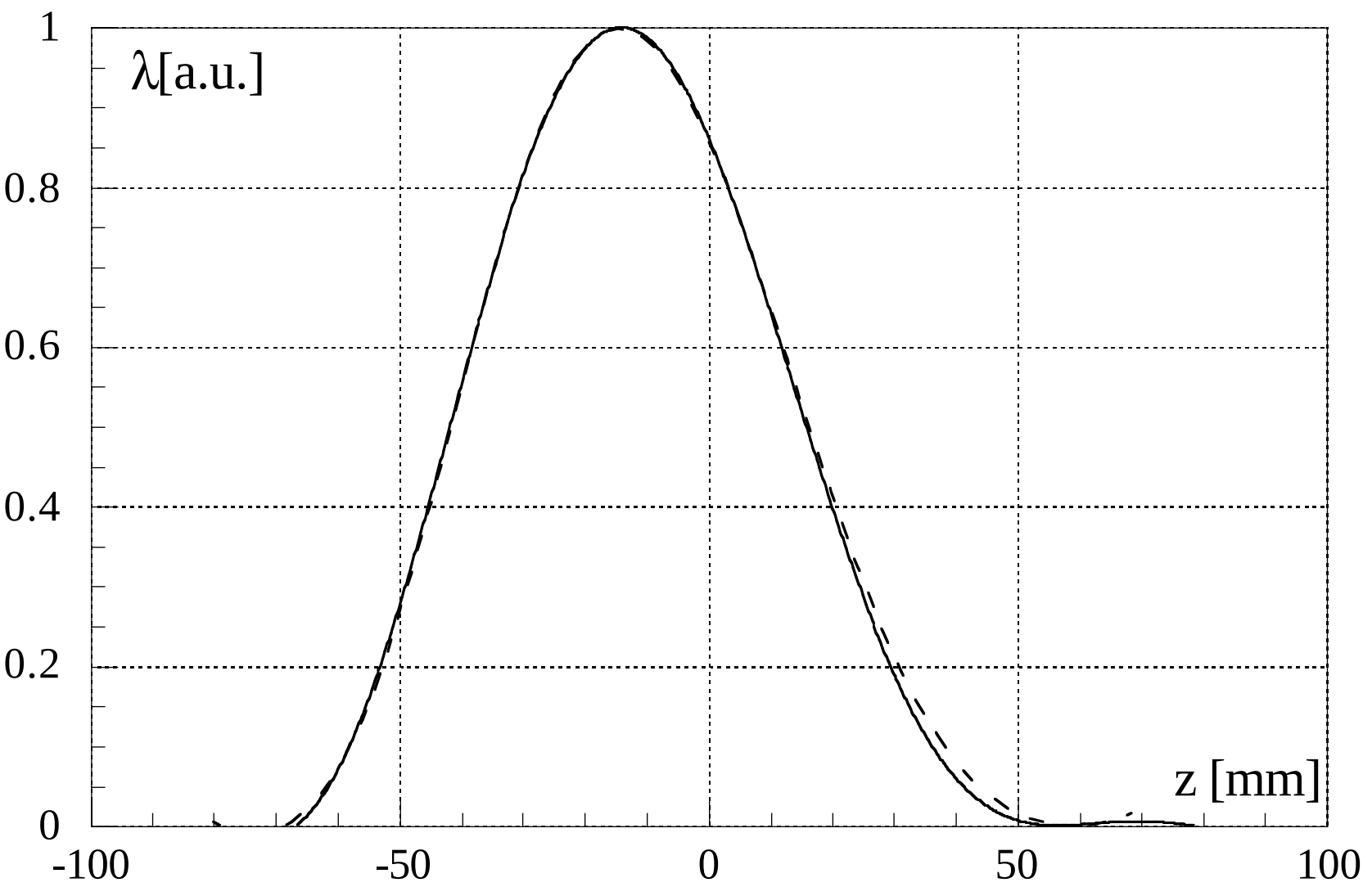

Figure 13: Bunch current distribution at 100 kV (I = 26 mA).
Solid line - measured signal; dotted line - numerical simulation.

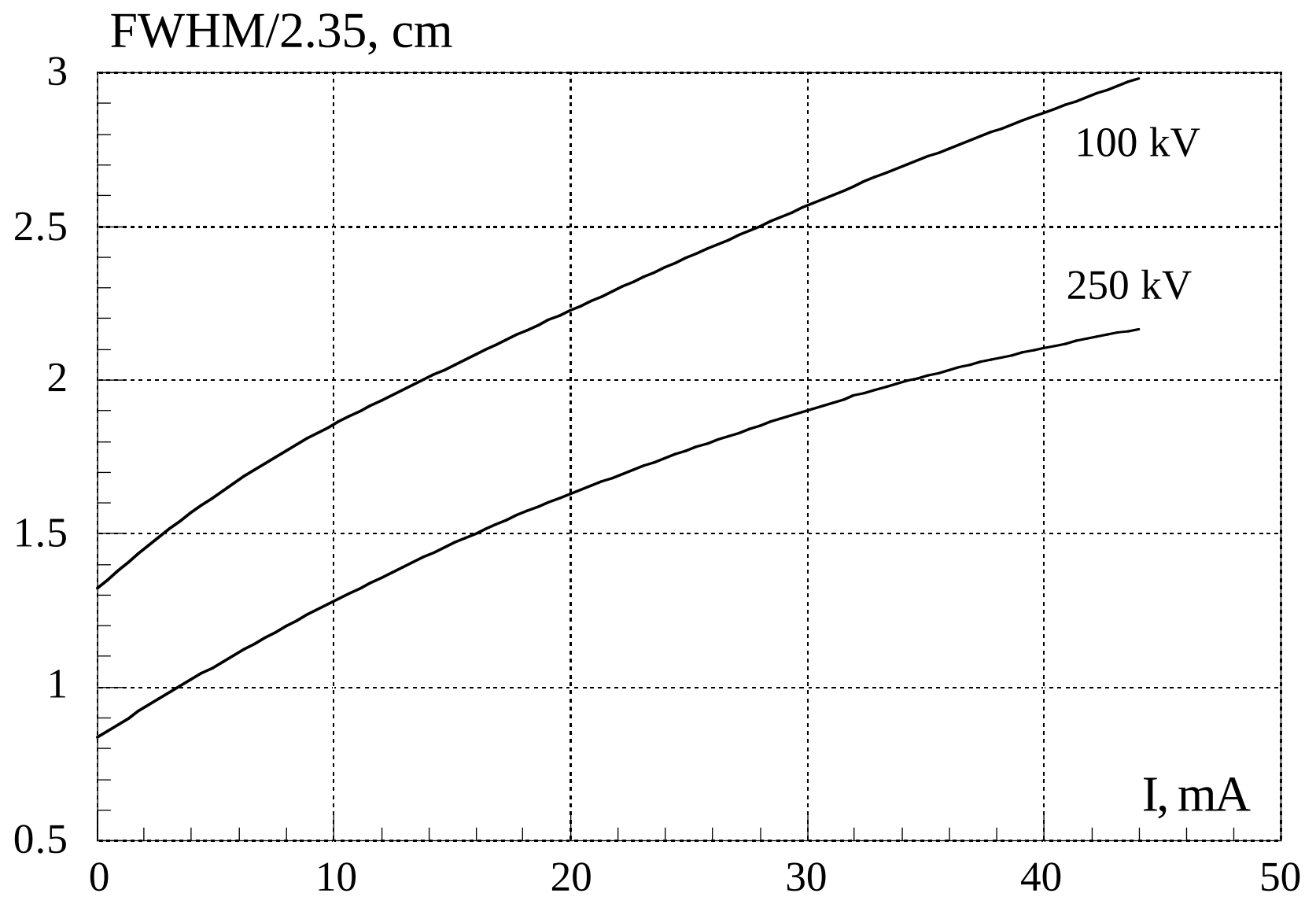

Figure 14: Bunch lengthening in DAΦNE at $V_{RF}$ = 100 kV and 250 kV.

## 5. Conclusions

The estimated machine normalized longitudinal coupling impedance |Z/n| is lower than 0.6 Ω in the frequency up to 20 GHz. The bunch length (FWHM/2.3548..) at nominal current can be varied from 2.2 cm at $V_{RF}$ = 250 kV to 3 cm at $V_{RF}$ = 100 kV.

Numerical simulations based on multiturn tracking of a large number of macroparticles interacting through the calculated numerically wake field (shown in Fig. 1) suit well to describe bunch lengthening and bunch shape in DAΦNE at different working conditions.

Double Water Bag model predictions of the longitudinal coherent mode coupling so far coincide with experimental observations. According to the model, the RF voltages higher than 150 kV would be preferable for DAΦNE operation in order to avoid the radial mode coupling of the lowest azimuthal modes (dipole, quadrupole and sextupole) at nominal current. Such mode coupling leads to the bunch shape modulations which could be harmful for beam-beam interactions.

## 7. Appendix

To investigate the single bunch behavior in the turbulent regime, we start from the Vlasov equation:

$$\frac{\partial\psi(z,\varepsilon;t)}{\partial t} = -c\frac{\partial\psi(z,\varepsilon;t)}{\partial\varepsilon}\frac{\partial H(z,\varepsilon;t)}{\partial z} + c\frac{\partial\psi(z,\varepsilon;t)}{\partial z}\frac{\partial H(z,\varepsilon;t)}{\partial\varepsilon} \qquad \text{(A.1)}$$

with $\psi(z,\varepsilon;t)$ the distribution function, and $H(z,\varepsilon;t)$ the single particle Hamiltonian. By linearizing the function $\psi(z,\varepsilon;t)$ around the stationary distribution and with the azimuthal mode expansion of the perturbed distribution function, we obtain for each azimuthal number $m$ the general equation:

$$\begin{aligned}\left[\Omega - m\omega(J)\right]R_m(J) = -i\frac{\alpha_c ce^2N}{4\pi^2\omega(J)ET_o}\frac{\partial\psi_o(J)}{\partial J}\int_0^{2\pi}d\phi\int_0^{2\pi}d\phi'\int_{-\infty}^{\infty}d\omega\int_0^{\infty}dJ'\\ R_l(J')\varepsilon(J,\phi)e^{i(m\phi-l\phi')}Z(\omega)e^{i\frac{\omega}{c}\left[z'(J'\phi')-z(J\phi)\right]}\end{aligned} \qquad \text{(A.2)}$$

where we have introduced the action and angle variables $J$ and $\phi$, with $\Omega$ the coherent oscillation frequency, $\omega(J)$ the synchrotron frequency depending on the oscillation amplitude, $R_m(J)$ the radial function of the $m^{th}$ azimuthal mode of the perturbed distribution, $\psi_o(J)$ the stationary distribution, and $Z(\omega)$ the longitudinal coupling impedance.

If we make the further assumption that the bunch is gaussian the above equation can be simplified to:

$$\begin{aligned}(\Omega - m\omega_{so})R_m(\hat{z}) = -i\frac{mce^2N}{T_o}\frac{1}{\hat{z}}\frac{\partial\psi_o(\hat{z})}{\hat{z}}\sum_{l=-\infty}^{\infty}i^{(m-l)}\\ \int_{-\infty}^{\infty}\frac{Z(\omega)}{\omega}J_m\left(\frac{\omega}{c}\hat{z}\right)d\omega\int_0^{\infty}R_l(\hat{z}')J_l\left(\frac{\omega}{c}\hat{z}'\right)\hat{z}'d\hat{z}'\end{aligned} \qquad \text{(A.3)}$$

with $\omega_{so}$ the synchrotron frequency, $\hat{z}$ the single particle amplitude of oscillation and $J_m(x)$ the Bessel function of the first kind of $m^{th}$ order.

By expanding the radial function $R_m(\hat{z})$ in terms of orthogonal polynomials, and considering only the most prominent radial mode, we obtain a simple eigenvalue equation:

$$(\Omega - m\omega_{so})a_m = \sum_{l=-\infty}^{\infty} M^{ml} a_l \tag{A.4}$$

to be satisfied for every mode number $m$, and with

$$M^{ml} = i\frac{\alpha_c c^2 e^2 N}{2\pi T_o E\omega_{so}\sigma_z^2}\frac{mi^{(m-l)}}{\sqrt{m!l!}}\int_{-\infty}^{\infty}\frac{Z(\omega)}{\omega}\left(\frac{\omega\sigma_z}{\sqrt{2}c}\right)^{m+l} e^{-\frac{\omega^2\sigma_z^2}{c^2}}d\omega \tag{A.5}$$

The coherent frequencies $\Omega$ are therefore obtained by solving the equation:

$$\det\left[\mathbf{M} - \mathbf{I}(\Omega - m\omega_{so})\right] = 0 \tag{A.6}$$

where $\mathbf{M}$ is the matrix given by equation (A.5), and $\mathbf{I}$ the identity matrix. The instability arises when the frequencies $\Omega$ become complex. This happens when two different longitudinal modes couple together. The longitudinal mode coupling theory could give a microwave threshold higher than the measured one.

For a complete analysis of the mode coupling, one should include in the treatment also the radial modes of oscillation for every azimuthal number. Unfortunately solutions for the Vlasov equation in this case can be found only when one considers a very simple distribution function, such as the so-called double water-bag distribution.

In the phase space it is described by the equation:

$$\psi(J) = \overline{\psi}\left[(1-\Gamma)U(J_1 - J) + \Gamma U(J_2 - J)\right] \tag{A.7}$$

where the constant $\overline{\psi}$ is derived from the normalization condition, $\Gamma$ is a parameter between 0 and 1 to better approximate the double water-bag to the real distribution, and $U(J)$ is the step function.

From the above equation, the radial modes are

$$R_m(J) = A_m \delta(J_1 - J) + B_m \delta(J_2 - J) \tag{A.8}$$

with $\delta(J)$ the symbolic Dirac delta function. By using the above relation in the Vlasov equation, we obtain the eigenvalue system:

$$\begin{cases} [\Omega - m\omega(J_1)]A_m = -(1-\Gamma) \sum_{l=-\infty}^{\infty} \left( M_{11}^{ml} A_l + M_{12}^{ml} B_l \right) \\ [\Omega - m\omega(J_2)]B_m = -\Gamma \sum_{l=-\infty}^{\infty} \left( M_{21}^{ml} A_l + M_{22}^{ml} B_l \right) \end{cases} \tag{A.9}$$

with

$$M_{ij}^{ml} = i \frac{\alpha_c e^2 N}{8\pi^3 T_o E[(1-\Gamma)J_1 + \Gamma J_2]\omega(J_i)} \int_0^{2\pi} \varepsilon(J_i,\phi) e^{-im\phi} d\phi \int_0^{2\pi} e^{il\phi'} w\left[z(J_i,\phi) - z(J_j,\phi')\right] d\phi' \tag{A.10}$$

and where $w(z)$ is the machine wake field.

The determinant of the eigenvalue system gives the eigenfrequencies $\Omega_{m,k}$. In such a way it is possible to take into account the coupling of radial modes with different azimuthal number.